# A Micropolar Cohesive Damage Model for Delamination of Composites

Md. M Rahaman[1], S P Deepu[1], D Roy[1,*] and J N Reddy[2]

[1]Computational Mechanics Lab, Department of Civil Engineering, Indian Institute of

Science, Bangalore 560012, India

[2]Advanced Computational Mechanics Lab, Department of Mechanical Engineering,

Texas A&M University, College Station, Texas 77843-3123

[*]Corresponding author; email: royd@civil.iisc.ernet.in

## Abstract

*A micropolar cohesive damage model for delamination of composites is proposed. The main idea is to embed micropolarity, which brings an additional layer of kinematics through the micro-rotation degrees of freedom within a continuum model to account for the micro-structural effects during delamination. The resulting cohesive model, describing the modified traction separation law, includes micro-rotational jumps in addition to displacement jumps across the interface. The incorporation of micro-rotation requires the model to be supplemented with physically relevant material length scale parameters, whose effects during delamination of modes I and II are brought forth using numerical simulations appropriately supported by experimental evidences.*

## 1. Introduction

Delamination is a commonly encountered mode of failure in laminated composite structures. Onset and propagation of delamination generally result in considerable reduction in the load carrying capacity, possibly triggering sudden structural collapse. Prediction of delamination initiation and its propagation have thus become topics of contemporary research interest, especially as composites are being extensively used in critically important aerospace structures and defense industries, among others. This has spawned a considerable literature that deals with theoretical modeling and experimental investigations, yielding a number of criteria for delamination initiation and propagation. However, only few of these are based on the physical mechanisms at the micro-structural level that cause inter-laminar fracture, even as one anticipates that an accurate prediction of delamination in real world applications should be woven around a physically consistent failure criterion. This is motivation enough for proposing a micromechanically founded delamination criterion incorporating intrinsic length scales and forms the aim of the present study.

A review of the literature reveals two existing approaches in modeling delamination. While the first one broadly works within the classical fracture mechanics setting, the second poses the problem as one in damage mechanics, softening plasticity, or a combination of the two [1]. The first approach, which employs classical fracture mechanics, uses stress-based criteria to predict delamination initiation [2, 3], and techniques based on linear elastic fracture mechanics (LEFM) such as virtual crack closure technique (VCCT) [4−8], J−integral method [9], virtual crack extension [10], or stiffness derivative [11] to model delamination propagation. However, finite element (FE) implementations of the LEFM techniques are fraught with difficulties, especially as the simulation of delamination growth may require complex moving mesh techniques [12]. Also, the calculation of fracture parameters makes use of the nodal variables as well as the topological information from the nodes behind and ahead of the crack front, computations of which are extremely cumbersome when a progressive crack growth is involved [13]. Many of these difficulties may be readily overcome if recourse is taken to the framework of damage mechanics. The concept of cohesive zone modeling (CZM), [14−16] is the most widely used interface damage model for the numerical simulation of delamination. The CZM relates the traction and displacement jump occurring at the interface between two layers. This model facilitates the integration of both delamination initiation and propagation. Decohesion elements provide appropriate criteria for the initiation and propagation of delamination without the prior knowledge of the crack location and propagation direction, thereby predicting the non-self-similar delamination growth [17]. Although, The FE implementation using decohesion elements is quite straightforward [18−23], it allows for a mesh-independent representation of material softening only with a very refined mesh [24]. Moreover, the FE analysis faces convergence issues when the interfacial strength is higher [26].

Many of these limitations could be overcome if mesh-free shape functions are used in lieu of the conventional FE bases [27], as they enable the introduction of a numerical length scale through the radius of the kernel used in the integral function representation. Useful though it is, such a model by itself does not include the intrinsic length scale parameter to reflect on the micromechanics of delamination. It would thus appear that there is a need to fall back on a lower scale cohesive zone modeling when the geometric length scale is smaller compared to the cohesive length scale [34].Clearly, then, a more accurate prediction of delamination is not ensured by the mere deployment of mesh-free shape functions alongside the traditional CZM.

The objective of this work is to develop a physically consistent micropolar cohesive damage model that could be used to predict, possibly with enhanced accuracy, delamination initiation and propagation. The organization of the rest of the paper goes as follows. Section 2 briefly describes the micropolar elasticity theory used in this work and the construction of the micropolar cohesive model for delamination of composites. Equations of equilibrium and their discretization are presented in Section 3. This is followed by numerical illustrations and concluding remarks in Sections 4 and 5, respectively.

## 2. Micropolar Model for Delamination

### 2.1 Basic Equations of Micropolar Elasticity:

In a micropolar continuum, besides the usual displacement vector field $u$, an additional field of micro-rotation vector $\varphi$ is introduced. This micro-rotation is different from macro-rotation, which is the curl of the displacement vector $u$. The introduction of micro-rotation results in an asymmetric strain tensor $\varepsilon$ and a micro-curvature tensor $\kappa$ (the latter also called the wryness tensor) given by (see [39])

$$\varepsilon_{ij} = \frac{\partial u_j}{\partial x_i} - e_{ijk}\varphi_k \tag{1}$$

$$\kappa_{ij} = \frac{\partial \varphi_j}{\partial x_i} \tag{2}$$

where $e_{ijk}$ denote components of the third order permutation tensor. The strain tensor $\varepsilon$ and the micro-curvature tensor $\kappa$ are work conjugates to the asymmetric stress tensor $\sigma$ and the couple stress tensor $\mu$ respectively; see [28–32, 39] for a more detailed exposition. The constitutive equations for linear micropolar elasticity are given as

$$\sigma_{ij} = D_{ijkl}\varepsilon_{kl} \tag{3}$$

$$\mu_{ij} = \Psi_{ijkl}\kappa_{kl} \tag{4}$$

For materials like composites, which are of current interest, the constitutive tensors **D** and $\Psi$ typically correspond to the anisotropic micropolar elasticity, an account of which may be found in Lesen [40]. It so happens that the anisotropy of composites modeled as a micropolar continuum may often be described as orthotropic for the conventional stress and isotropic for couple stress [38].

Delamination analysis may be performed based on a two-dimensional plane strain problem as suggested by Alfano and Crisfield [1]. Presently, the constitutive equations for the micropolar plane strain problem are chosen to be in the form:

$$\begin{bmatrix} \sigma_{11} \\ \sigma_{22} \\ \sigma_{12} \\ \sigma_{21} \\ \mu_{13} \\ \mu_{23} \end{bmatrix} = \begin{bmatrix} \dfrac{1-\nu_{32}\nu_{23}}{E_2 E_3 D_c} & \dfrac{\nu_{21}+\nu_{31}\nu_{23}}{E_2 E_3 D_c} & 0 & 0 & 0 & 0 \\ \dfrac{\nu_{12}+\nu_{13}\nu_{32}}{E_1 E_3 D_c} & \dfrac{1-\nu_{31}\nu_{13}}{E_1 E_3 D_c} & 0 & 0 & 0 & 0 \\ 0 & 0 & (G_{12}+G_m) & (G_{12}-G_m) & 0 & 0 \\ 0 & 0 & (G_{12}-G_m) & (G_{12}+G_m) & 0 & 0 \\ 0 & 0 & 0 & 0 & 2Gl^2 & 0 \\ 0 & 0 & 0 & 0 & 0 & 2Gl^2 \end{bmatrix} \begin{bmatrix} \varepsilon_{11} \\ \varepsilon_{22} \\ \varepsilon_{12} \\ \varepsilon_{21} \\ \kappa_{13} \\ \kappa_{23} \end{bmatrix} \tag{5}$$

where

$$G = \dfrac{E_1}{2(1+\nu_{12})} \tag{6}$$

$$D_c = \frac{1 - \nu_{12}\nu_{21} - \nu_{23}\nu_{32} - \nu_{13}\nu_{31} - 2\nu_{21}\nu_{32}\nu_{13}}{E_1 E_2 E_3} \tag{7}$$

$G_{12}$, $l$ and $G_m$ are respectively the shear modulus, internal length scale parameter for the laminate and micropolar shear modulus, with $\nu_{ij}$ ($i, j = 1, 2$ and $3$) denoting the Poisson's ratios. The symmetry of the constitutive matrix is ensured by using the reciprocal relations

$$\frac{E_{ij}}{\nu_i} = \frac{E_{ji}}{\nu_j} \qquad \left(\text{no sum over } i \text{ and } j\right) \tag{8}$$

Note that a plane stress anisotropic micropolar model can be obtained by the following changes in Eq. (5).

$$\frac{1 - \nu_{32}\nu_{23}}{E_2 E_3 D_c} \rightarrow \frac{E_1}{1 - \nu_{12}\nu_{21}}$$

$$(9)$$

$$\frac{\nu_{21} + \nu_{31}\nu_{23}}{E_2 E_3 D_c} \rightarrow \frac{\nu_{12} E_2}{1 - \nu_{12}\nu_{21}} \tag{10}$$

$$\frac{\nu_{12} + \nu_{13}\nu_{32}}{E_1 E_3 D_c} \rightarrow \frac{\nu_{21} E_1}{1 - \nu_{12}\nu_{21}} \tag{11}$$

$$\frac{1 - \nu_{31}\nu_{13}}{E_1 E_3 D_c} \rightarrow \frac{E_2}{1 - \nu_{12}\nu_{21}} \tag{12}$$

The length scale parameter $l$ in the constitutive model attempts at bridging the micro-mechanics with the macro-continuum by enabling the micro-rotation terms in the governing equations. One may observe that the rotational stiffness becomes smaller for smaller values of $l$ with the stress tensor regaining its symmetric nature when $G_m$ equals zero [30]. Thus one recovers the classical continuum as a limiting case of the micropolar theory.

## 2.2 Micropolar Cohesive Law:

At the interface where delamination is known to initiate and propagate, the classical traction separation law provides for the relevant constitutive equations by relating the cohesive surface traction, $\tau$ to the displacement jump, $\Delta$. This phenomenological model, also known as the cohesive law or the decohesion law, is popularly used to model the crack surfaces (see [13, 24, 17, 34] for a state-of-the-art on CZM). Over a period of time, Dugdale [14], Needleman [34], Rice and Wang [43], Tvergaard [41], Tvergaard and Hutchinson [42], Xu and Needleman [46], Camacho and Ortiz [44], Geubelle and Baylor [47] *et al*. have proposed several versions of the CZM, which are tabulated in Chandra *et al*. [45].

Of interest here is a modified traction separation law that accommodates the micropolar continuum. Accordingly, in addition to the usual stress tractions and displacement jumps, couple-stress tractions and rotational jumps must also be considered. The resulting CZM, which incorporates material length scale parameters, is referred to as the micropolar cohesive zone model (MCZM). In the micropolar traction separation law, Eq. (13) relating the stress traction and the displacement jump is supplemented with Eq. (14), which relates the couple traction $\tau_\theta$ with the rotation jump $\Delta_\theta$ through the intrinsic cohesive surface length scale ($l_c$)

$$\tau_i = K_p(1-D)\Delta_i \qquad i = n,\ t \tag{13}$$

$$\tau_\theta = K_p(1-D)l_c^{\ 2}\ \Delta_\theta \tag{14}$$

Here the suffixes $n$ and $t$ respectively denotes the normal and tangential components; $\theta$ represents the micro-rotation and $K_p$ the initial penalty stiffness and $D$ the scalar damage parameter. The following relation may be used to prevent inter-penetration of the crack faces.

$$\tau_n = K_p\Delta_n \quad \text{when } \Delta_n \le 0 \tag{15}$$

Now, equivalent traction $\tau_e$ and equivalent displacement jump $\lambda$ are defined respectively as

$$\tau_e = \sqrt{\tau_n^2 + \tau_t^2 + (\tau_\theta/l_c)^2} \tag{16}$$

$$\lambda = \sqrt{\Delta_n^2 + \Delta_t^2 + (l_c\Delta_\theta)^2} \tag{17}$$

Next, for the present work, a micropolar bilinear separation law is considered:

$$G_c = \frac{1}{2}\tau_m\lambda_f \tag{18}$$

$$\tau_m = K_p\lambda_0 \tag{19}$$

where $\tau_m$ and $G_c$ are the maximum interface strength and the critical energy release rate (per unit of the newly created surface) for fracture, respectively. The shaded area in Fig. 1 represents $G_c$ for a particular fracture mode.

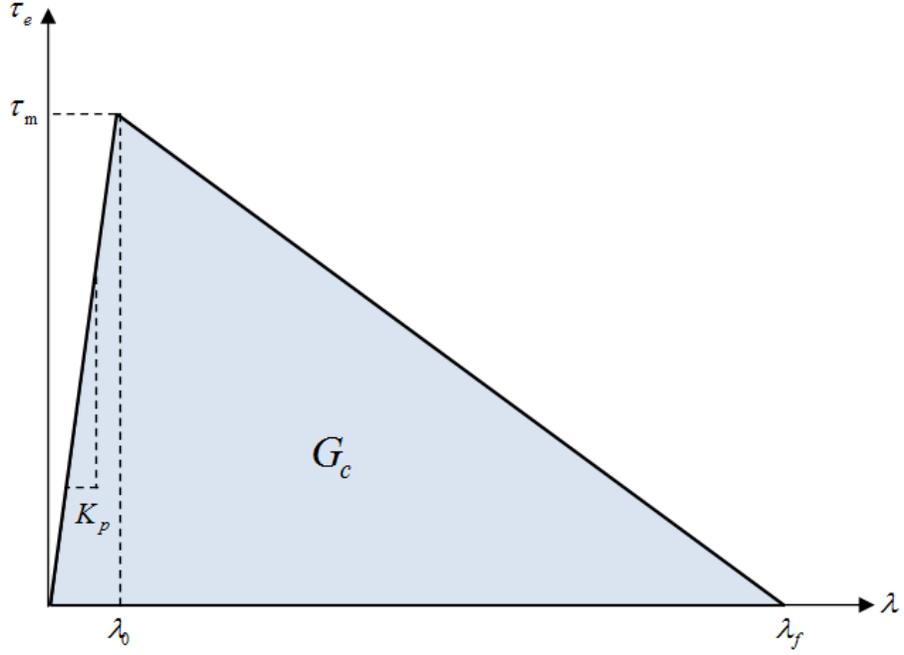

Fig. 1. Micropolar bilinear traction separation law

Once $\lambda_f$ and $\lambda_0$ are available respectively from Eq. (18) and Eq. (19), the damage variable $D$ may be computed as:

$$D(\lambda) = \begin{cases} 0 & \lambda \leq \lambda_0 \\ \dfrac{\lambda_f(\lambda - \lambda_0)}{\lambda(\lambda_f - \lambda_0)} & \lambda_0 < \lambda \leq \lambda_f \\ 1 & \lambda > \lambda_f \end{cases} \qquad (20)$$

## 3. Equations of Equilibrium and Discretization

Consider a 2D domain $\Omega$ that is split in two sub-domains $\Omega_1$ and $\Omega_2$ by a line of material discontinuity.

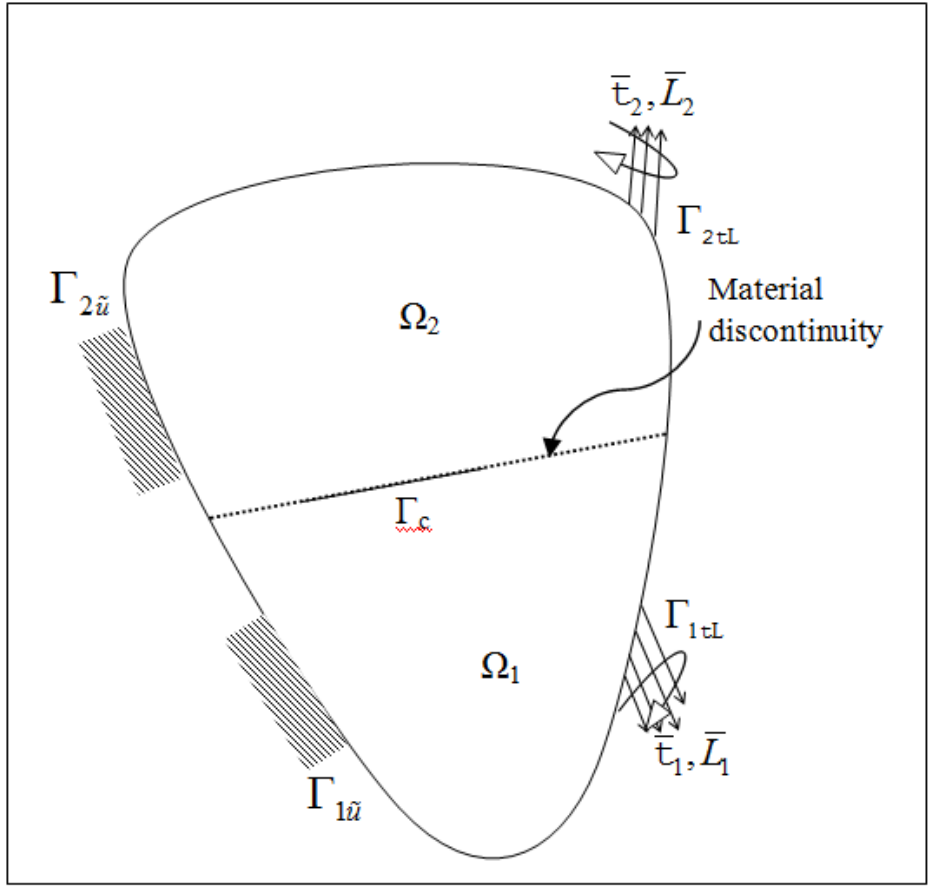

Fig. 2. Schematic of the problem description

Equilibrium equations for the micropolar continuum in the absence of body forces may be stated as follows:

$$\frac{\partial \sigma_{ij}}{\partial x_i} = 0 \quad \text{for } x \in \Omega \tag{21}$$

$$\frac{\partial \mu_{ij}}{\partial x_i} - e_{ijk}\sigma_{ik} = 0 \quad \text{for } x \in \Omega \tag{22}$$

$$\sigma_{ij}n_j = \overline{\mathbb{t}}_i, \ \mu_{ij}n_j = \overline{L}_i \quad \text{for } x \in \Gamma_{\mathbb{t}L}, \ \text{where } \Gamma_{\mathbb{t}L} = \Gamma_{1\mathbb{t}L} \cup \Gamma_{2\mathbb{t}L} \tag{23}$$

$$\tilde{u}_i = \overline{\tilde{u}}_i \quad \text{for } x \in \Gamma_{\tilde{u}}, \ \text{where } \Gamma_{\tilde{u}} = \Gamma_{1\tilde{u}} \cup \Gamma_{2\tilde{u}} \tag{24}$$

$$\sigma_{ij}n_j^c = \tau_i(\Delta_u) \tag{25}$$

$$\mu_{ij}n_j^c = T_i(\Delta_\varphi) \tag{26}$$

Here the generalized displacement vector $\tilde{u}$ contains both the displacement and the micro-rotation vectors. Moreover, $\overline{\mathbb{t}}$ and $\overline{L}$ are, respectively, the prescribed traction and couple at the traction boundaries $\Gamma_{1\mathbb{t}L}$ and $\Gamma_{2\mathbb{t}L}$, $\overline{\tilde{u}}$ is the prescribed generalized displacement at the essential boundaries $\Gamma_{1\tilde{u}}$ and $\Gamma_{2\tilde{u}}$, $n$ is the outward unit normal to parts of $\Gamma := \partial\Omega$ where tractions are prescribed, $n^c$ is the outward unit normal vector to the cohesive surface $\Gamma_c$, $\Delta_u$ is the displacement jump and $\Delta_\theta$ is the rotational jump across the line of material discontinuity, $\tau(\Delta_u)$ and $T(\Delta_\varphi)$ are respectively the stress and couple tractions developed at the interfacial boundary $\Gamma_c$ due to the displacement and rotational jumps.

The micropolar cohesive law is applied at $\Gamma_c$ via the duplicate node method (DNM) [37]. Specifically, two nodes are introduced at the same point on $\Gamma_c$ with one of them taken as part of $\Omega_1$ (domain 1) and the other as part of $\Omega_2$ (domain 2). The displacement and rotational jumps between these two nodes are determined, based on which the equivalent traction developed at the cohesive zone is calculated via the micropolar traction separation law.

The displacement jump $\Delta_u$ and rotational jump $\Delta_\theta$ are defined as follows:

$$\Delta_u(x) = u_1(x) - u_2(x) \qquad\qquad x \in \Gamma_c \qquad\qquad (27)$$

$$\Delta_\theta(x) = \varphi_1(x) - \varphi_2(x) \qquad\qquad x \in \Gamma_c \qquad\qquad (28)$$

$u_1(x)$ and $\varphi_1(x)$ are the displacement and rotation at a point $x$, when considered as $x \in \Omega_1$. Similarly, $u_2(x)$ and $\varphi_2(x)$ are the displacement and rotation at the same point $x$, when considered as belonging to $\Omega_2$. The outward direction is decided by whether the point is taken as part of $\Omega_1$ or $\Omega_2$. Discretization of the governing equations (21)-(26) using equations (27) and (28) lead to a system of nonlinear algebraic equations, which are solved through Newton's method. These details are given in Appendix A.

The problems considered in this work concern only mode I and mode II delaminations and the Reproducing Kernel Particle Method (RKPM) is used for domain/functional discretizations within a mesh-free setup. A brief account of RKPM shape functions (Aluru [35], Liu *et al.* [36], Shaw and Roy [25, 33]) is provided in Appendix B. Application of the DNM must tackle the issue of invertibility of the linearized problem (e.g. the stiffness matrix) as the duplicate nodes may not carry independent information, viz. when both the domains are assigned the same material properties. The resulting singularity may be removed by considering the interface as an internal boundary for each domain and applying, to the shape functions at the interface, corrections to impose the polynomial reproduction condition.

## 4. Numerical Illustrations

In this section, four numerical examples, each involving purely single-mode delamination, are considered to demonstrate the effect of the intrinsic length scale parameters (i.e., $l$ and $l_c$) that appear in the proposed model. The examples dealing with mode-I delamination are based on a couple of double-cantilever-beam (DCB) tests (Fig. 3), while those dealing with mode-II delamination relate to an end-loaded-split (ELS) test (Fig. 8) and an end-notched-flexure (ENF) test (Fig. 11). A variable vertical load is applied in the form of incremental displacements and the resulting non-linear equations are solved at each load step using Newton's update scheme. The initial delamination length $a_0$ (see Fig. 3, for instance) is imposed by setting the penalty stiffness $K_p$ to zero over this length. For all the reported simulations, $l_c$ and $G_m$ are chosen as 10 percent of $l$ and $G$ respectively.

### 4.1 Mode I Delamination

Numerical simulations of two different DCB tests are performed using both non-polar and micropolar RKPM schemes. A carbon fiber reinforced epoxy laminate (T300/977-2) with the elastic properties and geometry as given in Table 1 (where $B$ is the beam width) is considered in the first case. The second example, details of which are given in Table 2, is from Chen *et al.* [26]. For these two cases, the load-displacement curves are plotted in the form of the relative displacement between the two loading points versus the load applied. Figs. 4 and 6 show how the intrinsic length scale affects the delamination behavior for two different specimens. The simulation results through the proposed

MCZM are also compared with those based on the non-polar RKPM, FEM and the experimental data (see Figs. 5 and 7).

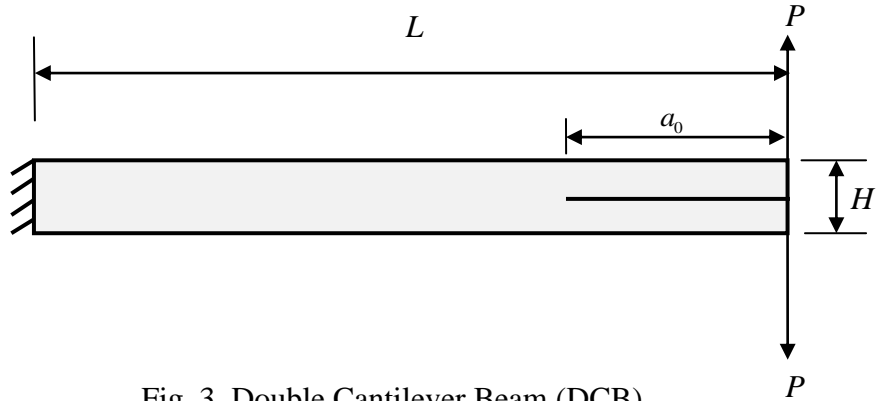

Fig. 3. Double Cantilever Beam (DCB)

**Table 1**

Dimensions and material properties for the DCB test T300/977-2 [24]

| $L$ (mm) | $H$ (mm) | $B$ (mm) | $a_0$ (mm) | $E_{11}$ (GPa) | $E_{22}$ (GPa) | $G_{12}$ (GPa) | $\nu_{12}$ | $\nu_{23}$ | $G_{IC}$ (N/mm) | $\tau_m$ (MPa) |
|---|---|---|---|---|---|---|---|---|---|---|
| 150 | 3.96 | 20 | 55 | 150 | 11 | 6 | 0.25 | 0.5 | 0.352 | 60 |

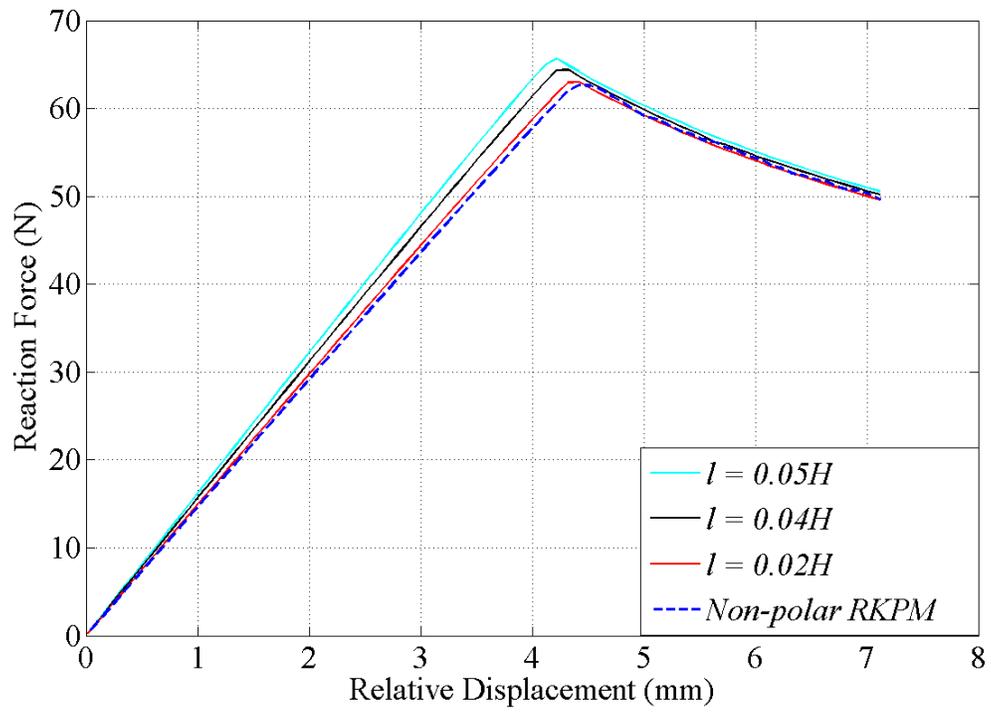

Fig. 4. Effect of length scale on load–displacement curve for DCB test T300/977-2

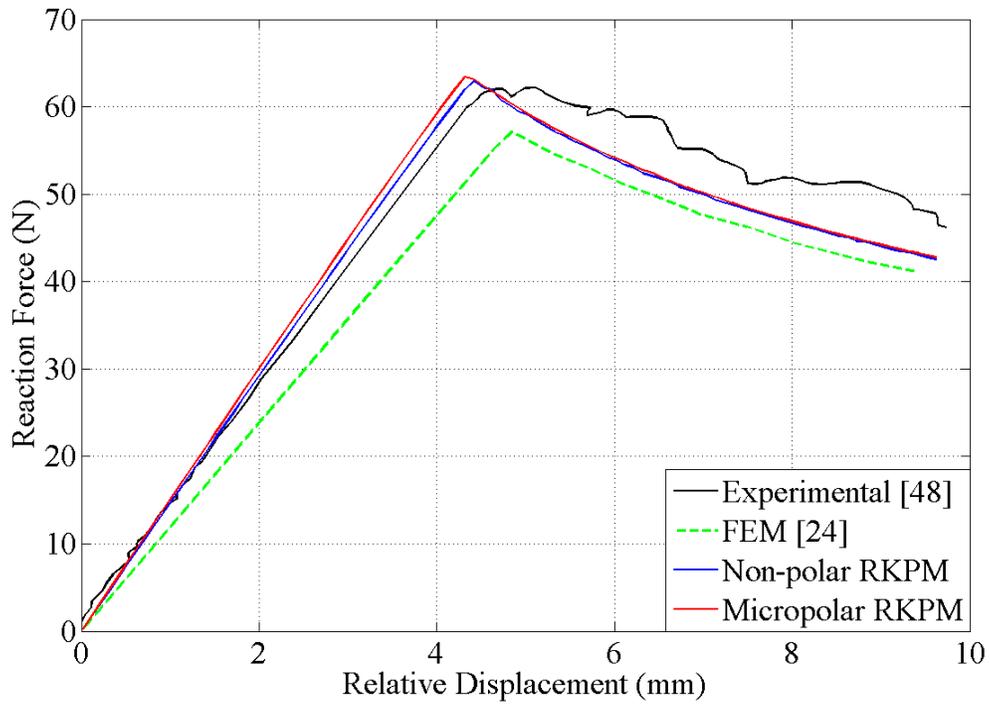

Fig. 5. Load–displacement curve for DCB test T300/977-2

**Table 2**

Dimensions and material properties for the second DCB test [26]

| $L$ (mm) | $H$ (mm) | $B$ (mm) | $a_0$ (mm) | $E_{11}$ (GPa) | $E_{22}$ (Gpa) | $G_{12}$ (GPa) | $\nu_{12}$ | $\nu_{23}$ | $G_{IC}$ (N/mm) | $\tau_m$ (MPa) |
|---|---|---|---|---|---|---|---|---|---|---|
| 150 | 3.1 | 2 | 22 | 130 | 8 | 6 | 0.25 | 0.45 | 0.257 | 48 |

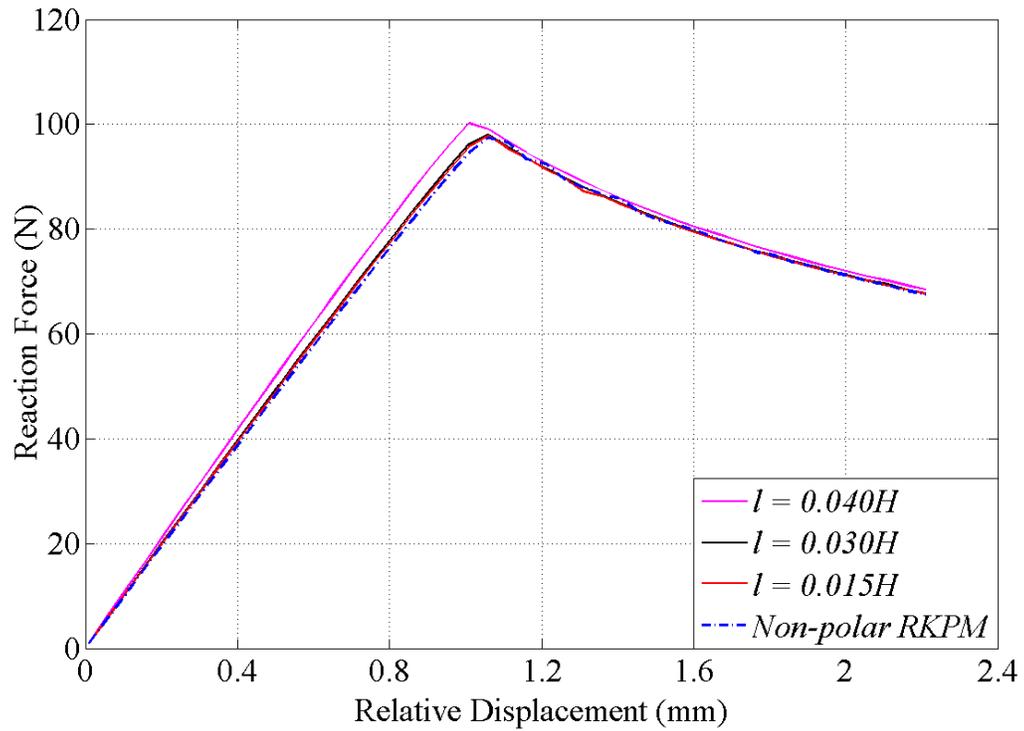

Fig. 6. Effect of length scale on load–displacement curve for second DCB test

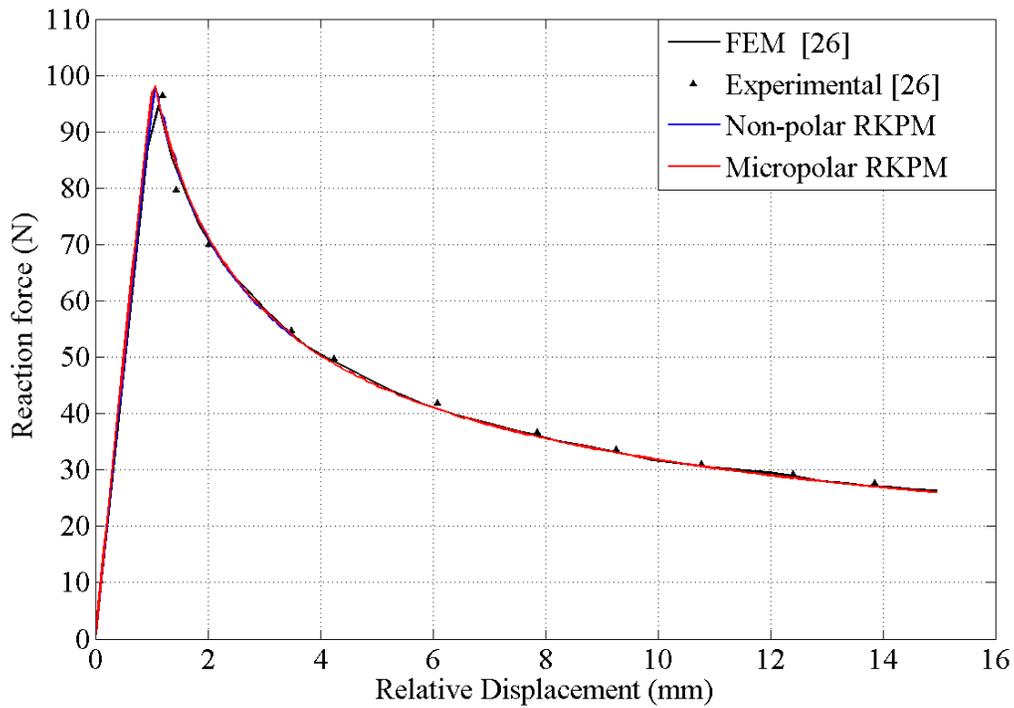

Fig. 7. Load – displacement curve for second DCB test

## 4.2 Mode II Delamination

Two cases of mode II delamination, namely, ELS and ENF, which are basically the in-plane sliding modes due to shear, are considered for numerical illustrations. For each of these test cases, simulations using both micropolar and non-polar RKPM are performed.

### 4.2.1 End Loaded Split Test

In this mode of delamination, the loading (in the form of incremental displacements) is applied at the free end of the bottom lamina. The geometric details and elastic properties given in Table 3 are used for numerical simulations of the ELS test.

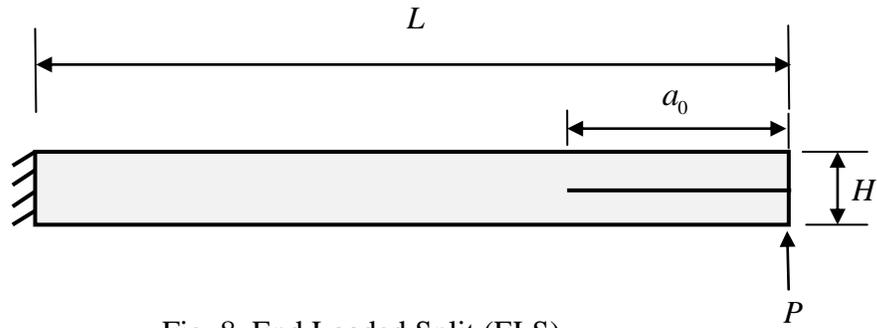

Fig. 8. End Loaded Split (ELS)

The effect of the intrinsic length scale parameter is shown in Fig. 9. The MCZM prediction shows a distinctively better agreement with the experimental result and this may be observed in Fig. 10.

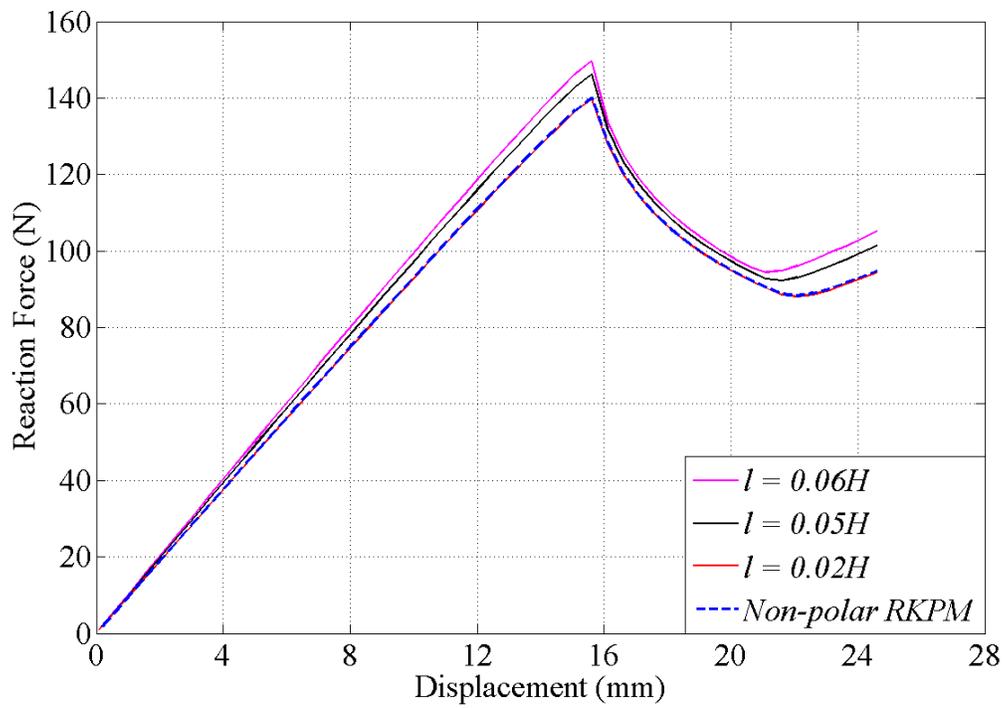

Fig. 9. Effect of intrinsic length scale on load–displacement curve for ELS test

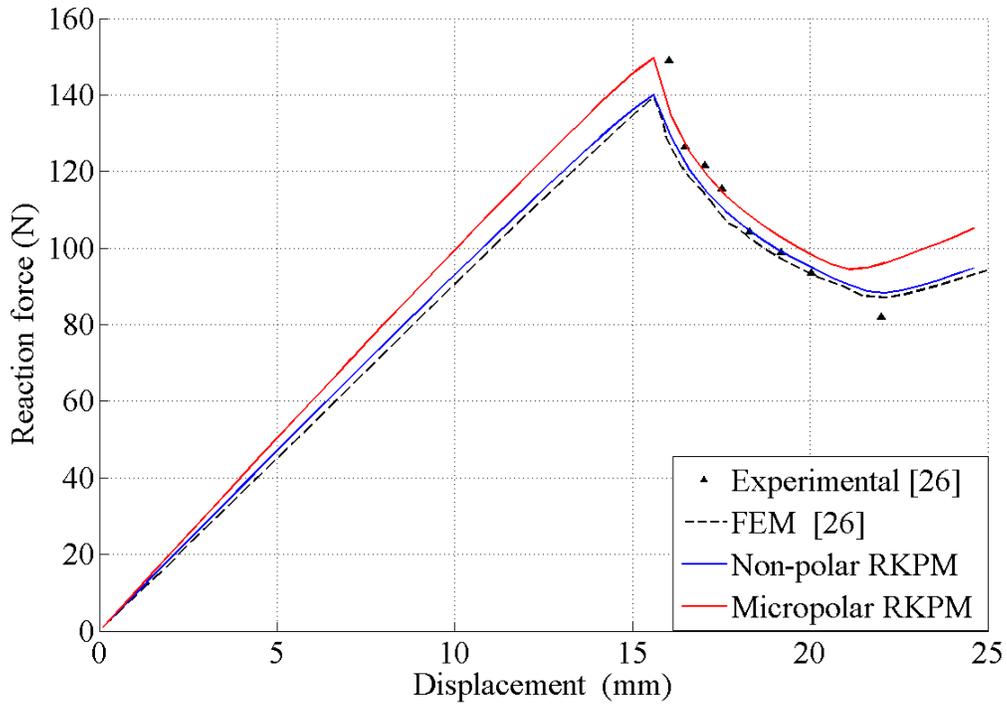

Fig. 10. Load−displacement curve for ELS test

**Table 3**

Dimensions and material properties for the ELS test [26]

| $L$ (mm) | $H$ (mm) | $B$ (mm) | $a_0$ (mm) | $E_{11}$ (GPa) | $E_{22}$ (GPa) | $G_{12}$ (GPa) | $\nu_{12}$ | $\nu_{23}$ | $G_{IIC}$ (N/mm) | $\tau_m$ (MPa) |
|---|---|---|---|---|---|---|---|---|---|---|
| 105 | 3.05 | 24 | 60 | 100 | 8 | 6 | 0.27 | 0.45 | 0.856 | 48 |

*4.2.2 End Notched Flexure Test*

An important sliding mode test is the ENF test in which a simply supported beam is subjected to a displacement controlled loading at the center of the beam with an initial delamination length $a_0$ (Fig. 11). Table 4 contains the geometric details and elastic properties of the ENF test specimen. Changes in the peak load with increase in the intrinsic length scale parameter are shown in Fig. 12. The MCZM predicts the critical displacement and peak reaction load quite accurately, albeit with some difference in the post peak behavior as depicted in Fig. 13.

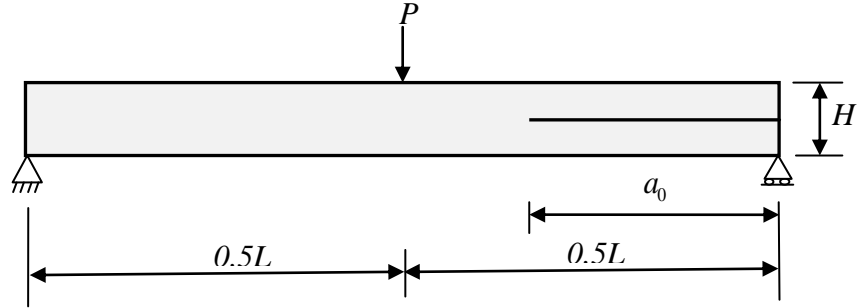

Fig. 11. End Notched Flexure (ENF)

**Table 4**

Dimensions and material properties for the ENF test [26]

| $L$ (mm) | $H$ (mm) | $B$ (mm) | $a_0$ (mm) | $E_{11}$ (GPa) | $E_{22}$ (GPa) | $G_{12}$ (GPa) | $\nu_{12}$ | $\nu_{23}$ | $G_{IIC}$ (N/mm) | $\tau_m$ (MPa) |
|---|---|---|---|---|---|---|---|---|---|---|
| 102 | 3.12 | 25.4 | 39.3 | 122.7 | 10.1 | 5.5 | 0.25 | 0.45 | 1.719 | 100 |

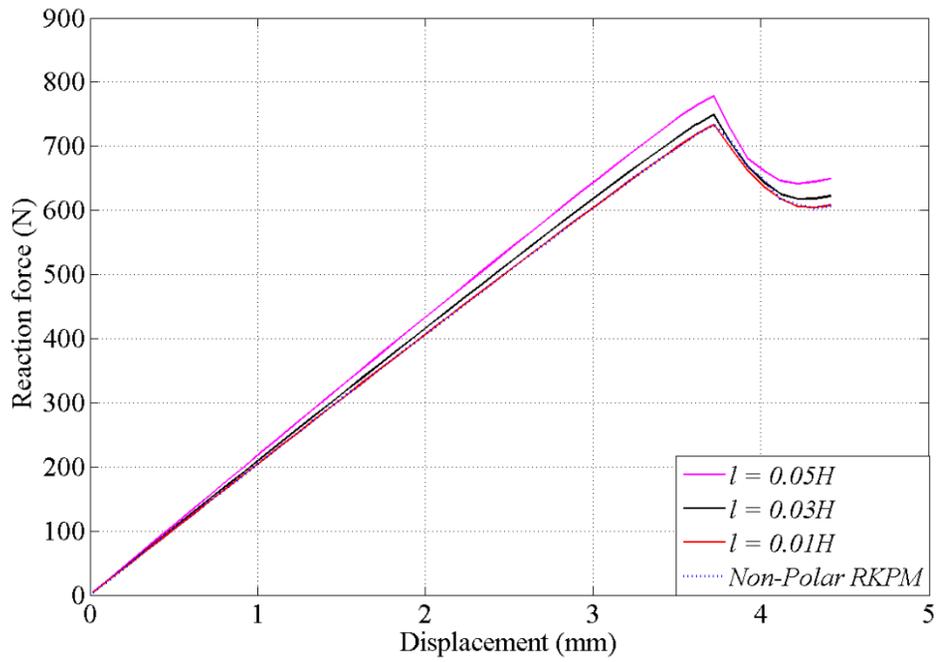

Fig. 12. Effect of length scale on load–displacement curve for ENF test

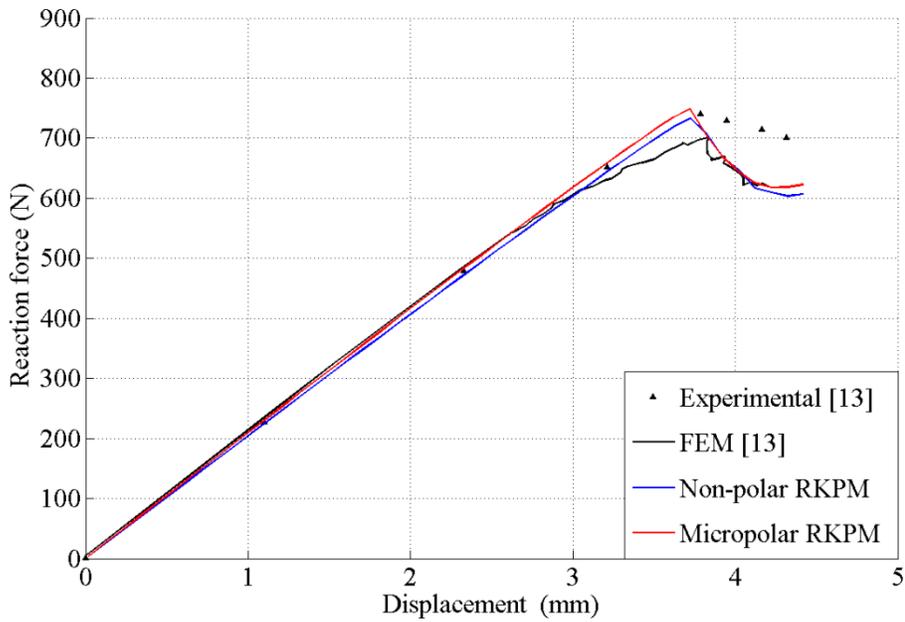

Fig. 13. Load – displacement curves for ENF test

All these numerical simulations indicate that both the micropolar and non-polar models predict the critical opening displacement for delamination reasonably well. However, in so far as the prediction of the peak load is concerned, non-polar models seem to err much more than the MCZM scheme in reproducing the experimental data. This of course would require choosing an appropriate length scale parameter, possibly by trial and error or, better still, based on the solution of an inverse problem [49]. Indeed, the micropolar cohesive damage model approaches its non-polar counterpart as the magnitude of the length scale parameter is decreased (see Figs. 4, 6, 9 and 12). In all the four test cases, as the magnitude of the length scale increases, the peak load in the load-displacement curve also increases. The effect of intrinsic length scales is more predominant in shear dominated modes vis-a-vis the normal mode.

## 5. Conclusions

A micropolar cohesive damage model is developed and applied to study delaminations in composite laminates. Use of micro-rotation degrees of freedom and the associated length scales needed in describing the macro-kinematics aim at bringing forth micromechanical effects within the continuum model. In the process, cohesive couple traction is considered alongside the standard cohesive forces at the interface of two different sub-domains. A penalty term, which is a function of the damage parameter, is varied along the interface so as to control the generalized traction such that, when fully damaged, the interface is rendered traction-free .The accuracy of the proposed model is illustrated through four numerical examples involving delamination of modes I and II. These illustrations establish that higher length scales are generally associated with higher peak loads and this may help reconcile better with the experimental evidence. The proposed model could be particularly useful for delamination studies on very thin laminates (e.g., thin films on substrates) as the micropolar effects may typically be more pronounced in such cases.

A more appropriate physical model for delamination should be provided by augmenting the continuum model with configurational force balances in the bulk and at the interface, whilst accounting for the surface energy. Such a perspective should render irrelevant the use of a cohesive law and would be the subject of a future study.

**Acknowledgement:** This work is funded by the Defense Research Development Organization, Government of India, through grant no. DRDO/0642.

## Appendix – A: Discretization and Inversion of the Governing Equations

Some details on the discretization and numerical solution, via linearization, of the governing equations for the micropolar model of delamination, considered in Section 2.2, are provided here. In a two-dimensional problem, the cohesive traction and couple traction are related to the displacement and rotation jumps as follows:

$$[\tau] = \begin{bmatrix} \tau_t \\ \tau_n \\ \tau_\theta \end{bmatrix} = \begin{bmatrix} K_t & 0 & 0 \\ 0 & K_n & 0 \\ 0 & 0 & K_\theta \end{bmatrix} \begin{bmatrix} \Delta_t \\ \Delta_n \\ \Delta_\theta \end{bmatrix} \tag{A.1}$$

where, $K_t = (1-D)K_p$

$$K_n = \begin{cases} (1-D)K_p & for \ \Delta_n > 0 \\ K_p & for \ \Delta_n \leq 0 \end{cases} \ and$$

$$K_\theta = (1-D)K_p l_c^2$$

The expression for the damage variable $D$, which is a function of the equivalent jump $\lambda(\Delta_t, \Delta_n, \Delta_\theta, l_c)$, is given in Eq. (20). The displacement and rotation jumps may be expressed as:

$$[\Delta] = \begin{bmatrix} \Delta_t \\ \Delta_n \\ \Delta_\theta \end{bmatrix} = \begin{bmatrix} N_{1c}^T U_1 - N_{2c}^T U_2 \\ N_{1c}^T V_1 - N_{2c}^T V_2 \\ N_{1c}^T \theta_1 - N_{2c}^T \theta_2 \end{bmatrix} = \tilde{N}_c^T R \tag{A.2}$$

where, $\tilde{N}_c^T = \begin{bmatrix} N_{1c}^T & 0 & 0 & -N_{2c}^T & 0 & 0 \\ 0 & N_{1c}^T & 0 & 0 & -N_{2c}^T & 0 \\ 0 & 0 & N_{1c}^T & 0 & 0 & -N_{2c}^T \end{bmatrix}$ and $R^T = [R_1^T \ R_2^T]$ . (A.3)

The displacement-rotation vector for $\Omega_j$ is $R_j^T = \begin{bmatrix} U_j^T & V_j^T & \theta_j^T \end{bmatrix}$ where $j = 1, 2$ . $N_1$ and $N_2$ are the RKPM shape functions belonging to the sub-domains $\Omega_1$ and $\Omega_2$, respectively. At the interface, $N_{1c} = N_1(X)$ and $N_{2c} = N_2(X)$ for $X \in \Gamma_c$. Thus $N_{1c}$ and $N_{2c}$ denote the shape functions for the duplicated nodes at $X \in \Gamma_c$. Discretization of the weak forms of equations (21)-(26) in Section 3 leads to:

$$\delta R_1^T K_1 R_1 + \delta R_2^T K_2 R_2 - \delta R_1^T F_{1tL} - \delta R_2^T F_{2tL} + \alpha \delta R_1^T S_1 R_1 - \alpha \delta R_1^T F_{1\tilde{u}} + \alpha \delta R_2^T S_2 R_2$$

$$-\alpha \delta R_2^T F_{2\tilde{u}} + \delta R_1^T (K_{11}^\delta R_1 - K_{12}^\delta R_2) + \delta R_2^T (-K_{12}^\delta R_1 + K_{22}^\delta R_2) = 0 \tag{A.4}$$

where,

$$K_i = \int_{\Omega_i} B_m^T E_m B_m \, d\Omega_i \qquad\qquad i = 1, 2 \tag{A.5}$$

$E_m$ is the constitutive matrix used in Eq. (5) and $B_m$ is the micropolar strain-displacement matrix relating the generalized strains (including the micropolar curvature components) with the generalized displacement $\tilde{u}$ (that includes the micro-rotation). $\alpha$ is a penalty parameter to impose the prescribed displacements and is typically assigned a value of order $10^9$. Moreover, we have (for $i, j = 1, 2$)

$$S_i = \int_{\Gamma_{i\tilde{u}}} N_i^T N_i \, d\Gamma_{i\tilde{u}} \tag{A.6}$$

$$F_{i\tilde{u}} = \int_{\Gamma_{i\tilde{u}}} N_i^T \bar{\bar{u}} \, d\Gamma_{i\tilde{u}} \tag{A.7}$$

$$F_{itL} = \int_{\Gamma_{itL}} N_i^T \bar{\bar{t}} \, d\Gamma_{itL} \tag{A.8}$$

where $\bar{\bar{t}}$ is the generalized prescribed traction, which includes both $\bar{t}$ and $\bar{L}$.

$$K_{ij}^\delta = \begin{bmatrix} K_{ij}^{\delta U} & 0 & 0 \\ 0 & K_{ij}^{\delta V} & 0 \\ 0 & 0 & K_{ij}^{\delta \theta} \end{bmatrix} \tag{A.9}$$

$$K_{ij}^{\delta U} = \int_{\Gamma_c} N_{ic}^T K_t(D) N_{jc} \, d\Gamma_c \tag{A.10}$$

$$K_{ij}^{\delta V} = \int\limits_{\Gamma_c} N_{ic}^T K_n(D) N_{jc}\, d\Gamma_c \tag{A.11}$$

$$K_{ij}^{\delta \theta} = \int\limits_{\Gamma_c} N_{ic}^T K_\theta(D) N_{jc}\, d\Gamma_c \tag{A.12}$$

Finally the following set of non-linear equations is arrived at:

$$[K + \alpha S + K^\delta(R)]R - F - \alpha F_{\underset{\sim}{u}} = f(R) = 0 \tag{A.13}$$

where,

$$K = \begin{bmatrix} K_1 & 0 \\ 0 & K_2 \end{bmatrix} \tag{A.14}$$

$$S = \begin{bmatrix} S_1 & 0 \\ 0 & S_2 \end{bmatrix} \tag{A.15}$$

$$F = \begin{bmatrix} F_{1\tilde{t}} \\ F_{2\tilde{t}} \end{bmatrix} \tag{A.16}$$

$$F_{\underset{\sim}{u}} = \begin{bmatrix} F_{1\tilde{u}} \\ F_{2\tilde{u}} \end{bmatrix} \tag{A.17}$$

$$K^\delta = \begin{bmatrix} K_{11}^\delta & K_{12}^\delta \\ K_{21}^\delta & K_{22}^\delta \end{bmatrix} \tag{A.18}$$

The non-linear Eq. (A.13) is solved through an iterative Newton update wherein the $(n+1)^{\text{th}}$ iteration at a load step is given by the recursive form:

$$R^{(n+1)} = R^{(n)} - (J^{(n)})^{-1} f(R^{(n)}) \tag{A.19}$$

$$J = \frac{\partial(f(R))}{\partial R} = K + \alpha S + \frac{\partial(K^\delta(R)R)}{\partial R} = K + \alpha S + K_T \tag{A.20}$$

The Jacobian matrix $J$ is obtained by linearization of Eq. (A.1).

$$\tau^{(n+1)} = \tau^{(n)} + \frac{\partial \tau}{\partial \Delta}(\Delta^{(n+1)} - \Delta^{(n)}) \tag{A.21}$$

where,

$$\frac{\partial \tau}{\partial \Delta} = \begin{bmatrix} \dfrac{\partial \tau_t}{\partial \Delta_t} & \dfrac{\partial \tau_t}{\partial \Delta_n} & \dfrac{\partial \tau_t}{\partial \Delta_\theta} \\[2ex] \dfrac{\partial \tau_n}{\partial \Delta_t} & \dfrac{\partial \tau_n}{\partial \Delta_n} & \dfrac{\partial \tau_n}{\partial \Delta_\theta} \\[2ex] \dfrac{\partial \tau_\theta}{\partial \Delta_t} & \dfrac{\partial \tau_\theta}{\partial \Delta_n} & \dfrac{\partial \tau_\theta}{\partial \Delta_\theta} \end{bmatrix} = \begin{bmatrix} K_t - \dfrac{\partial D}{\partial \Delta_t} K_p \Delta_t & -\dfrac{\partial D}{\partial \Delta_n} K_p \Delta_t & -\dfrac{\partial D}{\partial \Delta_\theta} K_p \Delta_t \\[2ex] -\dfrac{\partial D}{\partial \Delta_t} K_p \Delta_n & K_n - \dfrac{\partial D}{\partial \Delta_t} K_p \Delta_n & -\dfrac{\partial D}{\partial \Delta_\theta} K_p \Delta_n \\[2ex] -\dfrac{\partial D}{\partial \Delta_t} K_p l_c^2 \Delta_\theta & -\dfrac{\partial D}{\partial \Delta_n} K_p l_c^2 \Delta_\theta & K_\theta - \dfrac{\partial D}{\partial \Delta_\theta} K_p l_c^2 \Delta_\theta \end{bmatrix} \qquad (A.22)$$

Considering Eq. (A.21) and Eq. (A.3), the $K_T$ is obtained as,

$$K_T = \int\limits_{\Gamma_c} \tilde{N}_c^T \frac{\partial \tau}{\partial \Delta} \tilde{N}_c d\Gamma_c \qquad (A.23)$$

## **Appendix – B: RKPM Shape Functions**

An integral kernel approximation $g_a(x)$ to a given function $g(x)$, $x \in \Omega$, is given by:

$$g^a(x) = \int\limits_\Omega \overline{w_d}(x-s) \, g(s) \, ds \qquad (B.1)$$

where,

$$\overline{w_d}(x-s) = C(x,s) w_d(x-s) \qquad (B.2)$$

The corrected kernel function $\overline{w_d}(x-s)$ is taken as the product of the correction function $C(x,s)$ and the kernel function $w_d(x-s)$. The parameter $d$ is called the support radius or dilation parameter. A discrete approximation to Eq. (B.1) with $n_p$ particles (nodes) in the domain $\Omega$ is given as:

$$g^a(x) = \sum_{k=1}^{n_p} N_k(x) \, g(x_k) \qquad (B.3)$$

where $g(x_k)$ is the nodal value at $x_k$ and $N_k(x)$ is the RKPM shape function for the particle $k$:

$$N_k(x) = C(x - x_k) w_d(x - x_k) V_k \tag{B.4}$$

$V_k$ is a measure of the support domain around the particle $k$. In 2D domains, the kernel function $w_d$ is taken as the product of $w(\eta_1)$ and $w(\eta_2)$ where $\eta_1$ and $\eta_2$ are appropriately normalized scalar co-ordinates and the function $w(\eta_i)$ is presently given by the third order cubic spline:

$$w(\eta_i) = \begin{cases} \dfrac{2}{3} - 4\eta_i^2 + 4\eta_i^3 & 0 \leq \eta_i \leq \dfrac{1}{2} \\ \dfrac{4}{3} - 4\eta_i + 4\eta_i^2 - \dfrac{4}{3}\eta_i^3 & \dfrac{1}{2} < \eta_i \leq 1 \\ 0 & \eta_i > 1 \end{cases} \tag{B.5}$$

In this work, all the essential boundary conditions are enforced via the penalty method as the RKPM shape functions do not satisfy the Kronecker delta property.